\documentclass[
amsmath, amssymb, 
aps, pra,
reprint, twocolumn,
superscriptaddress,
longbibliography,
floatfix,
showkeys
]{revtex4-1}

\usepackage{graphicx}
\usepackage{csquotes}
\usepackage[urlcolor=blue, colorlinks=true,citecolor=blue]{hyperref}

\usepackage{color}
\usepackage{bbm}
\usepackage{nicefrac}

\definecolor{orange}{RGB}{255,127,0}

\def\avg#1{\mathinner{\langle{#1}\rangle}}
\def\bra#1{\ensuremath{\mathinner{\langle{#1}|}}}
\def\ket#1{\ensuremath{\mathinner{|{#1}\rangle}}}

\newcommand{\tr}{\text{Tr}}

\DeclareMathOperator{\Var}{Var}

\input{Qcircuit}

\begin{document}
\title{Decoding quantum errors with subspace expansions}

\author{Jarrod R. McClean}
	\email{Corresponding author: jmcclean@google.com}
	\affiliation{Google Inc., 340 Main Street, Venice, CA 90291, USA}
\author{Zhang Jiang}
	\affiliation{Google Inc., 340 Main Street, Venice, CA 90291, USA}
\author{Nicholas C. Rubin}
	\affiliation{Google Inc., 340 Main Street, Venice, CA 90291, USA}
	\author{Ryan Babbush}
	\affiliation{Google Inc., 340 Main Street, Venice, CA 90291, USA}
\author{Hartmut Neven}
	\affiliation{Google Inc., 340 Main Street, Venice, CA 90291, USA}
\date{\today}     

\begin{abstract}
With the rapid developments in quantum hardware comes a push towards the first practical applications on these devices.  While fully fault-tolerant quantum computers may still be years away, one may ask if there exist intermediate forms of error correction or mitigation that might enable practical applications before then.  In this work, we consider the idea of post-processing error decoders using existing quantum codes, which are capable of mitigating errors on encoded logical qubits using classical post-processing with no complicated syndrome measurements or additional qubits beyond those used for the logical qubits.  This greatly simplifies the experimental exploration of quantum codes on near-term devices, removing the need for locality of syndromes or fast feed-forward, allowing one to study performance aspects of codes on real devices.  We provide a general construction equipped with a simple stochastic sampling scheme that does not depend explicitly on a number of terms that we extend to approximate projectors within a subspace.  This theory then allows one to generalize to the correction of some logical errors in the code space, correction of some physical unencoded Hamiltonians without engineered symmetries, and corrections derived from approximate symmetries.  In this work, we develop the theory of the method and demonstrate it on a simple example with the perfect $[[5,1,3]]$ code, which exhibits a pseudo-threshold of $p \approx 0.50$ under a single qubit depolarizing channel applied to all qubits.  We also provide a demonstration under the application of a logical operation and performance on an unencoded hydrogen molecule, which exhibits a significant improvement over the entire range of possible errors incurred under a depolarizing channel.
\end{abstract}

\maketitle

\section{Introduction}
Since the inception of the idea by Feynman~\cite{Feynman:1982}, considerable progress has been made in the understanding and implementation of quantum computation.  However, while a key development in the road map of quantum computing was the concept of quantum error correction, the hardware requirements to implement fully fault-tolerant schemes for non-trivial algorithms may still be some years away.  A natural question that arises from this realization is whether it will be possible to perform meaningful computations on non-fault tolerant or noisy intermediate scale quantum computers (NISQ)~\cite{Preskill:2018}.  Experimental and theoretical proposals have explored the potential for performing a well-defined computational task faster than a classical computer on as few as 50 qubits, a task often referred to as ``quantum supremacy''~\cite{Boixo:2016,Neill195}.  It remains an open question, however, if these results can be extended to applications of interest outside the domain of pure computation.

Many early proposals for practical applications have advocated the use of variational algorithms~\cite{Peruzzo:2014,McClean:2016Theory,OMalley:2016,Farhi:2014,Kandala:2017,Shen2017Quantum,Santagati:2018,Dumitrescu:2018,Li:2017td,Colless:2018,Romero:2018,Hempel:2018,McClean2018,farhi2018classification}, which are known to experience a natural form of robustness against certain types of noise.  In conjunction with this, much progress has been made in reducing the gate overhead required for practical applications, especially in the domain of quantum chemistry~\cite{Motta:2018,Kivlichan:2018,Babbush:2018,Babbush:2018b}.  However, the impact of incoherent noise remains daunting for the accuracy thresholds specified~\cite{OMalley:2016,Peruzzo:2014,Sawaya:2016}.  As a result, in order to reach practical applications, it may be necessary to implement some form of partial error correction for NISQ computations.  The exact form this error correction could take to achieve success is yet unknown; however, it has been suggested that one of the best applications for early quantum computers is using them to study and optimize error correcting codes in real conditions~\cite{Iyer:2018}.  Yet despite great theoretical progress, most quantum codes are difficult to study experimentally on NISQ devices due to the need for complicated syndrome measurements, fast feedback, and decoding capabilities.

An alternative approach that strays from traditional ideas of error correction and targets NISQ devices is ``error mitigation''.  This term largely refers to techniques that reduce the influence of noise on a result using only batch measurements and offline classical processing as opposed to active measurement and fast feedback type corrections.  While they are not believed to lead to scalable, fault-tolerant computation, it is hoped that sufficient mitigation may open the possibility of practical applications or inspire more near-term error correction ideas.  A number of these techniques have been developed both within an application specific and general context ~\cite{McClean:2017hybrid,Endo:2017,Temme2017error,Otten:2018}.  If one specializes to the quantum structure of fermionic problems, notably the N-representability conditions, enforcing these as constraints alone can reduce the impact of noise in simulations~\cite{Rubin:2018}.  More generally within quantum simulation, an error mitigation technique known as the quantum subspace expansion (QSE)~\cite{McClean:2017hybrid} was predicted and experimentally confirmed to both approximate excited states and reduce errors through additional measurements and the solution of a small offline eigenvalue problem~\cite{Colless:2018}.  Since then there have been variations leveraging QSE that use both additional techniques from quantum chemistry for excited states~\cite{Parrish:2019} and imaginary time evolution~\cite{Motta:2019}.

In this work, we show that it is possible both to use existing quantum error correcting codes to mitigate errors on NISQ devices and to study the performance of these codes under experimental conditions using classical post-processing and additional measurements. We briefly review the theory of stabilizer codes~\cite{gottesman1997stabilizer} and post-processing in this framework, which we then generalize using quantum subspace expansions.  While a connection to symmetries was explored in the original work~\cite{McClean:2017hybrid} and this connection was extended in subsequent work~\cite{Bonet:2018,McArdle:2018} that has also been verified by experimental implementation~\cite{Sagastizabal:2019}, these papers have focused on application specific contexts.  Here we generalize this to any circuit performed within a quantum code, and show how subspace expansions may be used to then correct some logical errors within the code space, as well as be applied to approximate symmetries of unencoded Hamiltonians. We provide a concrete example using the perfect $[[5,1,3]]$ code to demonstrate post-processed quantum state recovery.  When applied at the highest level, this recovery exhibits a $p \approx 0.50$ pseudo-threshold for an uncorrelated depolarizing channel applied to all qubits.  An example of an unencoded hydrogen molecule is also demonstrated across the entire range of depolarizing errors.  We close with an outlook and potential applications of this methodology.

\begin{figure}[t]
\centering
\includegraphics{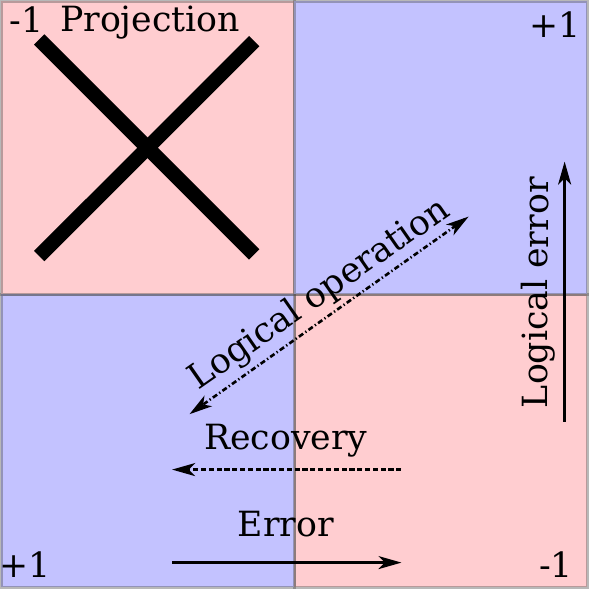}
    \caption{Cartoon schematic of error correction vs error projection in a stabilizer code.  We sketch the quantum space as divided into the blue code space, defined by $+1$ eigenvalues of stabilizers and the red non-code space as defined by having $-1$ eigenvalues for some of the stabilizers.  In traditional error correction, the stabilizers are measured, the errors decoded, and recovery operations are applied to return one to the code space.  In error projection, we use projectors based on stabilizers to remove sections of non-code space using only simple Pauli measurements and post-processing.  We can also combine this technique with forms of recovery, but the effective difference is depicted by the discarding of large parts of errant Hilbert space.
    \label{fig:ProjectionSchematic}}
\end{figure}

\section{Correcting logical observables in post-processing}
We begin by briefly reviewing and establishing notation for the relevant topics of quantum error correcting codes in the stabilizer formalism, and using this formalism to develop a set of projection operators.  Consider a set of $n$ physical qubits.  Quantum error correcting codes utilize entanglement to encode a set of $k < n$ logical qubits, with the hope of improving robustness to probable errors.  A code that requires at least a weight $d$ Pauli operator to induce a logical error is said to have distance $d$.  These three numbers are often used to define a quantum error correcting code, with the notation $[[n, k, d]]$. 

One of the most prominent classes of quantum error correcting codes are stabilizer error correction codes.  In this formalism, the stabilizer group is a commutative group with a set of generators $S \subset \mathcal{P}_n$, where $\mathcal{P}_n$ is the Pauli group on $n$ qubits and the logical states are defined to be the simultaneous $+1$ eigenstates of the stabilizer group $S$ generated by the set of stabilizer generators $S$.  The set of $2k$ logical operators formally written $\mathcal{L} = \{\overline X_i, \overline Z_i\}_{i=1,\ldots,k}$ perform the desired Pauli operation on states in the code space.  A related formulation that we will use here, is that the code space of logical states is defined by the degenerate ground state space of the code Hamiltonian
\begin{align}
    \mathcal{H}_c = -\sum_{M_i \in \mathcal{M}} M_i
\end{align}
where $\mathcal{M}$ is a set of check operators drawn from the stabilizer group that can be used to deduce error syndromes.  More explicitly, $S \subseteq \mathcal{M} \subseteq \mathcal{S}$, such that the minimal set is the stabilizer generators, but additional operators from the stabilizer group may be added, as in techniques where one uses redundancy in stabilizer operators to ameliorate the need for multiple measurements passes, also called single shot error correction~\cite{Bombin:2015,Campbell:2019}.

Traditional error correction proceeds by measuring the check operators $M_i$, typically using some ancillary qubits, and using the resulting syndrome information to decode and recover from the error.  Considering a simple example, if $\ket{\overline \Psi}$ is a state in the code space, and an error that is correctable within the given code takes one from the code space to a state outside of it, denoted $\ket{\Psi_e}$, a recovery map takes $\ket{\Psi_e} \rightarrow \ket{\overline \Psi}$ before proceeding further with the computation.  However, if we restrict ourselves to a NISQ device which cannot perform on-the-fly measurement, decoding, and recovery, then at the end of the error channel, we expect in this simple example that the state will be modeled by a mixed state density matrix $\rho_e = (1-p) \ket{\overline \Psi} \bra{\overline \Psi} + p \ket{\Psi_e}\bra{\Psi_e}$.  In this example with the assumption one cannot perform the requisite syndrome measurement, it's clear there is no unitary recovery operation that can remove the error.  One must use projection operators to remove the errors from the state in this situation. 

Fortunately, the stabilizer formalism suggests such a set of projectors.  Namely, members of the stabilizer group, $M_i \in \mathcal{S}$ have eigenvalues $\pm 1$ and may each be used to construct a projector $P_i = (I + M_i)/2$ that removes components of the state outside the $+1$ eigenspace of the stabilizers or code space. One may use this to construct a projector that is a linear combination of projectors to remove desired errors outside the code space.  We note that this clearly cannot remove logical errors made within the code space.  This is related to the idea of error detection and post selection which has made recent progress both in theory and experiment~\cite{gottesman2016quantum,li2017fault,linke2017fault,trout2018simulating,harper2019fault,vuillot2017error,rosenblum2018fault,willsch2018testing}. Quantum error detection typically discards results based on syndrome measurements without using correction, however we will avoid the need for direct syndrome measurements, which can be cumbersome on geometrically local qubit layouts and are challenging to do in a fault tolerant fashion for complex codes.  For a stabilizer group with generators $S_i$, the complete projector can be formed from $\prod_{S_i \in S} (I+S_i)/2 = \prod_i P_i$.  We note that due to the commutative structure of the generators, that these projectors commute and hence their products are also projectors.  It is also the case that if $S_i S_j = S_k$, then $P_i P_j P_k = P_i P_j$ and hence only projectors associated with the generators are required for the complete projection.  When taken over all the generators, the expression $\prod_i P_i$ is the sum of all elements of the stabilizer group with a constant coefficient which we fix to $1/2^m$,
\begin{align}
    \overline P = \overline P^\dag = \prod_i^m P_i = \frac{1}{2^m} \sum_{M_i \in \mathcal{S}} M_i
\end{align}
where $m$ is the number of stabilizer generators used.  For the case of full projection, this will be the full stabilizer group which contains $2^m$ terms.  While this is generally an exponential number of terms, it will be shown that the number of terms is not an explicit factor in the cost when a stochastic sampling scheme is used to apply the corrections.  Rather the correction cost will depend on the volume of the state outside the code space.  The group structure allows projective correction of the density matrix $\overline P \rho \overline P^\dagger$ on a NISQ device to be relatively straightforward.  As all the $M_i$ are simple Pauli operators, it is relatively simple to evaluate the projectively corrected value of a physical observable $\Gamma$.  Suppose that some logical Hermitian operator $\Gamma$ is expressed as a sum of Pauli operators $\Gamma_i$ as $\Gamma = \sum_i \gamma_i \Gamma_i$.  Then the corrected value of $\Gamma$ may be computed from
\begin{align}
    \avg{\Gamma} &= \frac{1}{c} \tr \left[\overline P \rho \overline P^\dagger \Gamma \right] = \frac{1}{c} \tr \left[\rho \left(\overline P^\dagger \Gamma \overline P \right)\right] \notag \\
    &= \frac{1}{c 2^{2m}} \sum_{ijk} \gamma_j \tr \left[\rho M_i^\dagger \Gamma_j M_k\right] \notag \\
    &= \frac{1}{c 2^{m}} \sum_{jk} \gamma_j \tr \left[\rho \Gamma_j M_k\right] \\
    c &= \tr \left[\overline{P} \rho \overline{P}^\dagger \right] = \tr \left[\overline{P} \rho \right]
\end{align}
where we have used that for these commuting, Hermitian projectors, $\overline P = \overline P^\dagger = \overline P^\dagger \overline P = \overline P \overline P^\dagger$, and logical operators $\Gamma$ commute with stabilizer group elements $M_i$, and if $M_i^\dagger M_k$ is in the set of operators, we can rewrite it as a single sum over these operators which will be repeated.  In the case that we use the operators built from the stabilizer generator projectors here, this will always be the case.

As this expansion may contain a large number of terms, it is important to develop a scheme for sampling from it that maximizes efficiency.  One should not simply run down the list of terms and measure each one to a fixed precision, as this will have poor scaling.  Rather one should use a method that reflects the fact that if the state $\rho$ were completely contained in the code space, the measurement of $c$ should be $1$ and have $0$ variance.  This means that a reasonable stochastic sampling of terms should converge quickly, be somewhat independent of the actual number of terms, and depend on the quality of the state $\rho$.  We discuss a simple stochastic scheme for sampling these corrections and the associated cost of doing so in Section \ref{sec:random}.

We emphasize a distinction between this measurement scheme and traditional error correction/detection is that we do not need to measure the stabilizers in earnest.  As this is a post-processing procedure, we are free to destroy the information in the state by measuring qubit-wise across Pauli operators.  To be explicit, if one had the Pauli operator $X_1 Z_2 Z_3 X_4$ as a stabilizer, a true stabilizer measurement would require extracting only the $\pm 1$ measurement using an ancilla.  However in this scheme, we are free to use repeated preparations of the state and construct any unbiased estimator of $\langle X_1 Z_2 Z_3 X_4 \rangle$ we desire, including those which might destroy the encoded state. This dramatically simplifies the use of codes with non-local stabilizer measurements.  

\section{Corrections with recovery operations}
\label{sec:recovery_operations}

The power of error correction extends beyond the simple identification of errors and includes recovery operations that restore some states to the original code space.  The formalism here built on projectors and post-processing would seem at first glance unable to take advantage of such unitary projection operations; however, we will show how one can use these recovery operations to some advantage in sampling complexity over the unrecovered projections.  

Consider a set of Pauli errors on the system of physical qubits $\{E_i\}$ which is known to be correctable within the chosen code.  These errors will either commute or anti-commute with the stabilizers of the code to produce a syndrome of the error that has happened, which we denote $s_j^i$ for the $j$'th syndrome measurement of the $i$'th error.  We will assume the recovery operation for this error within the code is known, and is denoted as $R_i$.

The formalism presented here avoids direct stabilizer measurement by design to favor implementation on NISQ devices, hence we need to specify how one uses recovery operations within the projection formalism.  Similar to a projector on the code space, we may formulate a projector onto the error subspace that corresponds to error $E_i$ acting on the code space.  This is given by
\begin{align}
    \overline P_{E_i} = \prod_j \frac{1}{2}(I + (-1)^{s_j^i} S_j)
\end{align}
where $s_j^i \in \{0, 1\}$ is the syndrome associated with the error $E_i$ and stabilizer generator $S_j$.  Once one has projected into this space, we can now use the recovery operation, $R_i$ to map the state back into the code space before using it.  If we take the set of all correctable errors, including no error as the identity, then we get an updated correction formula for projection with recovery as
\begin{align}
    \avg{\Gamma} & = \frac{1}{c} \sum_i \tr \left[  R_i \overline P_{E_i} \rho \overline P_{E_i}^\dagger R_i^\dagger \Gamma \right] \\
    c & = \sum_i \tr \left[\overline P_{E_i} \rho \right]
\end{align}
where now we have assumed the ability to apply the recovery operations $R_i$, however in many cases this again reduces to a simple sum over Pauli operators that may be stochastically sampled, where many of the same simplifications resulting from commutation of logical operators with stabilizers are possible.

The consequences of including recovery on top of projection are interesting.  The immediate practical benefit in increasing the size of Hilbert space over which one attains signal.  This is reflected in the estimation of the value $c$, and leads in practice to lower errors with small, finite samples.  However a tradeoff is being made in including these values in that it may reduce the overall projection quality.  Consider for example a distance $d$ code.  If one has a Pauli error with weight greater than $(d - 1)/ 2$, a recovery operation may become a logical error which is not then removed by this procedure.  In contrast, strict projection is capable of removing errors of all the way up to weight $d-1$, which is a significant boost in maximum potential.  However, as mentioned, the tradeoff of finite sampling complexity with potential for correction must be carefully balanced in real implementations.

\section{Relaxing projectors to subspace expansions}
In the previous section, we showed how explicit projectors from quantum error correcting codes can be used to correct observables in post processing.  Here we show how these constructions can be relaxed for greater flexibility and power with simple relations to approximations of these projectors within a subspace.  We know that for an expansion built from a product of projection based on the stabilizer generators, the coefficients may be chosen to be uniform.  However when one truncates terms from this series, this is no longer the case and we must consider a more general expression
\begin{align}
    \overline{P}_c = \sum_i^{2^m} c_i M_i
\end{align}
where the check operators still come from the stabilizer group, however it no longer needs to be true that $\overline{P} \propto \prod_i(I+S_i)$.
To find coefficients $c_i$, we formulate this problem as minimizing the distance to the code space subject to a normalization constraint.  Using the Hamiltonian formulation of the code space, this is equivalent to approximating the ground state of the code space by
\begin{align}
    & \min_{c_i} \tr \left[ \overline P_c \rho \overline P^\dagger_c \mathcal{H}_c \right] \notag \\
    & \text{such that } \tr \left[ \overline P_c \rho \overline P^\dagger_c \right] = 1 \notag \\
    &\overline P_c = \sum_i c_i M_i.
\end{align}
This optimization is dependent both on the state $\rho$ and choice of $\mathcal{H}_c$ in general.  This problem has a well-known exact solution, and is given by the solution of the generalized eigenvalue problem
\begin{align}
    HC &= SCE \\
    H_{ij} &= \tr \left[M_i^\dagger \mathcal{H}_c M_j \rho \right] \\
    S_{ij} &= \tr \left[ M_i^\dagger M_j \rho \right].
\end{align}
Where $H$ here forms a representation of the action of the code Hamiltonian in this stabilizer projector basis, the matrix $S$ is the overlap or metric matrix defining the subspace geometry, $C$ is the matrix of eigenvectors, and $E$ is the diagonal matrix of eigenvalues.  We note that for cases where the check operators $M_i$ are built from projectors from generators, the solutions coincide with the previous formalism.

If we denote the number of check operators used to define the subspace as $N_M$, there appears to be $O(N_M^2)$ matrix elements to measure here. However, we note that if the check operators used are from the stabilizer group as before (we relax this condition in subsequent sections), then we again have the property that $[M_i, \mathcal{H}_c] = 0$, and we can reduce the measurement $\tr \left[M_i^\dagger \mathcal{H}_c M_j \rho \right]$ to $\tr \left[\mathcal{H}_c M_k \rho \right]$ for $M_k = M_i^\dagger M_j$, where each $M_k$ can be easily precomputed.  This implies the number of matrix elements used in the estimate is actually linear in the number of check operators used.  As before, we prescribe a stochastic sampling method for each of the individual elements, and leave open potential optimizations that work on directly sampling the most important matrix elements applied to sample vectors as well.

When it is not the case that one builds the check operators from the product of stabilizer projectors, it provides an optimal solution that interpolates between different numbers of projectors in the subspace given. This type of expansion about a state is referred to as a quantum subspace expansion (QSE).  The ground state eigenvector of this $N_M \times N_M$ eigenvalue problem, forms the optimal solution of the above problem within this subspace.  We note as a technical detail the matrix $S$ may be singular due to lack of errors.  For this reason the generalized eigenvalue problem must be solved by canonical diagonalization, where the matrix $S$ is first diagonalized, eigenvectors associated with $0$ eigenvalues are discarded, and the problem is diagonalized in the resulting basis. 

So far we have exploited the properties of stabilizer code Hamiltonians and QSE to formulate a recovery procedure from non-code space errors. As in typical quantum error correction, the degeneracy of the ground state of the full code Hamiltonian prevents referencing a single state within the code space.  This makes removing logical errors with the above procedure impossible. However, when considered in conjunction with a problem Hamiltonian such as that from a quantum physical system like an electronic system, it becomes possible to correct logical errors as well if the goal is to prepare an eigenstate of this Hamiltonian or minimize its energy. 

If we denote the problem Hamiltonian expressed in the basis of logical operators (i.e. $X_i\rightarrow \overline X_i, Z_i \rightarrow \overline Z_i$) as $\overline{\mathcal{H}}_p$, this problem Hamiltonian has the effect of breaking the degeneracy of the ground state space of the code Hamiltonian.  If we now expand to a set of operators $\left\{ O_i \right\} \subseteq \text{Span}\left(\mathcal{S} \bigcup \mathcal{L} \right)$ and perform the above procedure on $\mathcal{H}_c + \overline{\mathcal{H}}_p$, then we can correct error within the logical space as well.  We note that in connection to the subsequent section on correcting unencoded Hamiltonians, symmetries of the problem Hamiltonian encoded in the logical space may be used as additional symmetry projector generators.  Simply summing the Hamiltonians together, however, is akin to a penalty method in constrained optimization and would require balancing of the contribution from the code and problem Hamiltonian. A more stable procedure would be to use the above QSE procedure to find corrected matrix elements in the basis of logical operators.  Then perform a subsequent diagonalization on $\overline{\mathcal{H}}_p$. We note that including logical operators when using the code Hamiltonian exclusively only serves to introduce additional errors due to mixing in the code space.  It is only through the breaking of the degeneracy through the encoded problem Hamiltonian that it becomes beneficial to include operators from the logical space as well.  This inclusion of logical operators also allows access to excited states of the problem Hamiltonian as in the original QSE work.

\section{Corrections with unencoded systems}
So far we have considered the case of decoding within an error correcting code that redundantly encodes quantum information via engineered symmetries.  However, this strategy inevitably involves some overhead due to the encoding in the execution of gates, and in some near-term experiments, it will still be most practical to work directly in the space of a physical problem Hamiltonian $H_p$.  Here we show how the machinery developed so far can be applied to this case.

In this case, the physical problem Hamiltonian $H_p$ may have  symmetries that are often known about the desired state ahead of time. For example in the case of an interacting fermion system, the total number of fermions, the total spin and $S_z$ component, and symmetries related to spatial degrees of freedom in a system are often good candidates.  The application of these symmetries has been explored previously, symmetries~\cite{McClean:2017hybrid,Bonet:2018,Bravyi:2017,Sagastizabal:2019}, and as these are expected to be exact symmetries, it is always safe to apply them when the symmetry is known.  

While these symmetries are exact and effective to apply, they are often more expensive to implement than one might desire.  For example, the problem of number symmetry in a fermion Hamiltonian can take eigenvalues that range from $0$ to the number of spin orbitals in the system.  Thus to select just the correct particle number, one may have to construct a projector which removes all the components except the desired particle number $N_p$, or $\propto \prod_{n \neq N_p} (n - \hat N)$ where $\hat N$ is the number operator on all the fermionic modes of the system.  As one can see, this may result in an unreasonable number of terms.  

As a result, it is much simpler and effective to start with symmetries that have only two distinct eigenvalues, also referred to as $\mathbb{Z}_2$ symmetries of the problem Hamiltonian.  An example of this is the number symmetry operator, which in the Jordan-Wigner representation takes the simple form $\prod_i Z_i$.  An extension of this, is to use both the up-spin ($\alpha)$ and down-spin $(\beta)$ number parities, which generate the full number parity, and offer additional power in their projection.  These are simply given by $\prod_{i \in \alpha} Z_i$ and $\prod_{i \in \beta} Z_i$ respectively.  These simple parity symmetries have been utilized before to reduce the number of qubits~\cite{Bravyi:2017}, however just as in subsystem error correcting codes, retaining these redundancies can sometimes be beneficial for gate depth or efficiency of representation.  That work also contained a general algorithm for searching for unknown $\mathbb{Z}_2$ symmetries in these Hamiltonians that can be used here.

One interesting advantage of the subspace expansion approach is that if one identifies a symmetry, but does not know the symmetry subspace to which the desired state will belong, it will be selected automatically.  Consider for example the case where we find a $\mathbb{Z}_2$ symmetry, $F$, of the Hamiltonian $H_p$, defined by $[H_p, F] = 0$, but we do not know which of the two eigenspaces the exact ground state belongs to.  In that case applying $P_F^{+}=(I + F)$ and $P_F^{-}=(I - F)$ can lead to drastically different results, and concatenations with other symmetries can compound this problem.  However the QSE procedure can automatically select between the two to find the optimal choice.

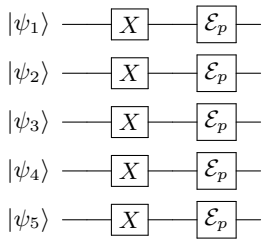
\begin{figure}[t]
\centering
\mbox{
\Qcircuit @C=1.0em @R=0.5em{
\lstick{\ket{\psi_1}}  & \qw & \gate{X} & \qw & \gate{\mathcal{E}_{p}} & \qw \\
\lstick{\ket{\psi_2}}  & \qw & \gate{X} & \qw & \gate{\mathcal{E}_{p}} & \qw \\
\lstick{\ket{\psi_3}}  & \qw & \gate{X} & \qw & \gate{\mathcal{E}_{p}} & \qw \\
\lstick{\ket{\psi_4}}  & \qw & \gate{X} & \qw & \gate{\mathcal{E}_{p}} & \qw \\
\lstick{\ket{\psi_5}}  & \qw & \gate{X} & \qw & \gate{\mathcal{E}_{p}} & \qw}}
\caption{Circuit used for simulating a noisy application of a transversal $\overline X$.  The single qubit depolarizing channel with probability $p$, $\mathcal{E}_p$, is implemented to mimic the effect of an imperfect gate.  The 5 physical qubits of the $[[5,1,3]]$ code are labeled as $\ket{\psi_i}$. \label{fig:transversal_circuit}}
\end{figure}

One may also use this to consider beneficial projectors that are derived from approximate symmetries of the Hamiltonian or more generally operators that do not commute with the Hamiltonian, but have known structure with regards to the problem.  Consider for example the local fermion occupation operators $a_i^\dagger a_i$, which under the Jordan-Wigner transformation is given by $Z_i$.  In general, we expect $[H_p, Z_i] \neq 0$, and for the exact state to have some component on orbital $i$.  However, the sites in fermionic simulation problems are often approximately well ordered in terms of both energy and likely occupation in the so-called natural orbital basis.  This means that some sites are less likely to be occupied than others, and they can inflict disproportionately large energetic errors compared to similar errors on other sites. The QSE procedure can automatically decide whether to apply the projector $(I + Z_i)$ by balancing the contribution of site $i$ to the exact wavefunction against the energetic damage done by its extra occupation under noise in this example.  While this particular projector is simple enough that it is tantamount to a removal of a qubit when applied exactly, one can imagine pair occupation projectors on the highest energy orbitals, such as $(I + Z_i Z_j)$ can be effective in removing errors that caused by erroneous occupation of the highest energy orbitals which cannot be mitigated by simple truncation of a qubit.  While we do not explore these explicitly in our examples, they are a fruitful area of future development.

\section{Example demonstrations}
Here we both exhibit some of the performance of the presented techniques and clarify their construction through the use of simple examples.  Both a general error correcting code and specific problem Hamiltonian systems with symmetries are studied.  We note that in our numerical studies we use a single qubit depolarizing channel defined by
\begin{align}
    \mathcal{E}_p(\rho) = \left( 1 - p \right) \rho + \frac{p}{3} \left( X \rho X + Y \rho Y + Z \rho Z \right)
\end{align}
which corresponds to the convention that the totally mixed state is achieved at $p=3/4$.  A different convention is sometimes used in the literature that corresponds to the totally mixed state at $p=1$, and the two can be converted simply.  

\subsection{$[[5,1,3]]$ code}
To see how the general recovery process using stabilizer codes can work in practice, let us consider the concrete example of the perfect $[[5,1,3]]$ code, which is a distance $3$ code that encodes $1$ logical qubit in $5$ physical qubits.  This code has logical operators and stabilizer generators 
\begin{align}
    \overline{X} =& XXXXX \\
    \overline{Z} =& ZZZZZ \\
    S =& \{XZZXI, IXZZX, \notag \\
    & \ XIXZZ, ZXIXZ\}.
\end{align}
We denote the two states of the logical qubit as $\ket{\overline{0}}$ and $\ket{\overline{1}}$, and an arbitrary code space state that is a superposition of these two states as $\ket{\overline{\Psi}}$ or the pure state density matrix $\overline{\rho}$.  Suppose a single qubit error $E_a$ occurs on a code space state with probability $p$.  This yields a density matrix $\rho = (1-p) \ket{\overline{\Psi}}\bra{\overline{\Psi}} + p E_a\ket{\overline{\Psi}}\bra{\overline{\Psi}}E_a^\dagger$.  As this code has distance 3, at least one of the stabilizers anti-commutes with this error and we call this $S_a$.  If we use $S_a$ as an expansion operator, then we know
\begin{align}
    (I + S_a) E_a \ket{\overline{\Psi}} & = E_a \ket{\overline{\Psi}} - E_a S_a \ket{\overline{\Psi}} = 0 
\end{align}
which holds only from the property of $\ket{\overline{\Psi}}$ being in the code space and the stabilizer $S_a$ anti-commuting with the error $E_a$.  As a result any coefficient of the expansion procedure will remove this error from the density matrix, however we must still be careful to evaluate the trace.  As we are operating in a post-processing regime however, we do not generally expect a single isolated error to occur.  Rather, we expect many errors to occur and we want to see how our procedure treats this.  

We denote the hierarchy of check operators as the elements in the sum generated by $S^{(l)}=\prod_{i=1}^l (I + S_i)$, where the ordering of stabilizer generators has been fixed.  To see how this hierarchy works in practice, consider an uncorrelated depolarizing channel acting on all $5$ physical qubits with probability $p$.  In this situation, we have up to $5$ qubit errors, which we do not expect the code can recover from without introduction of a problem Hamiltonian, however they occur with probability $p^5$, which can be quite small for modest $p$.  

\begin{figure}[t!]
\centering
\includegraphics[width=8cm]{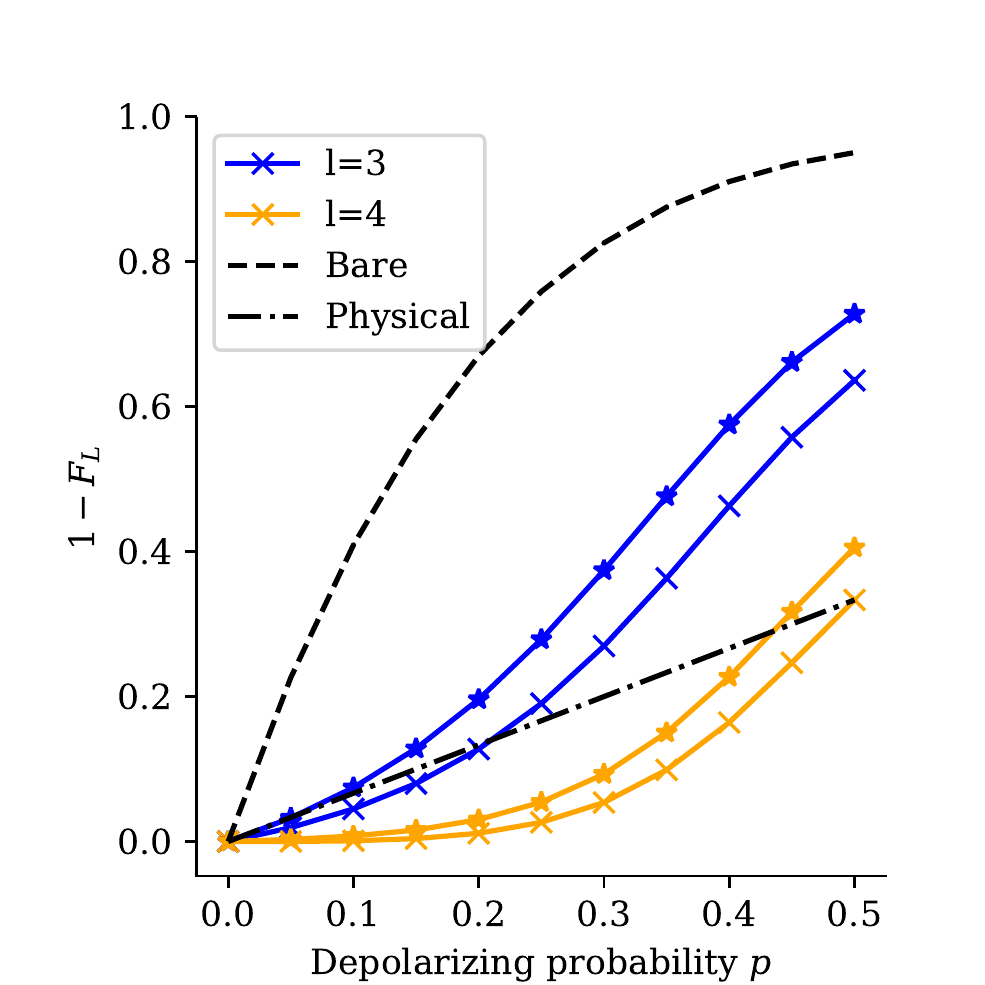}
    \caption{Pseudo-threshold crossover for recovery using the $[[5,1,3]]$ code under an uncorrelated depolarizing channel showing a $p \approx 0.50$ pseudo threshold for the full correction procedure.  We plot the logical infidelity $1-F_L$ where $F_L$ is the logical fidelity of a selected state in the code space of the $[[5,1,3]]$ code as a function of the depolarizing probability for each qubit $p$.  The label $l$ denotes the number of products from the stabilizer generators used in the expansion operator set. The physical line depicts the same error if the logical state is encoded in a single qubit in the standard way.  The ``bare'' line indicates the logical error rate with no recovery procedure applied.  The starred lines close to each level of the hierarchy show an approximation to that level of projection using the QSE projection with $2$ less check operators to demonstrate the smooth performance of the subspace procedure.
    \label{fig:PseudoThreshold}}
\end{figure}

To evaluate the performance in practice, we perform the following numerical experiment.  A logical state $\ket{\overline{\Psi}}$ in the $[[5, 1, 3]]$ code is prepared, then subjected to an uncorrelated depolarizing channel on all qubits with probability $p$. In connection with the formalism above, we evaluate the expectation value of the logical operator $A = \ket{\overline{\Psi}}\bra{\overline{\Psi}}$.  This operator does not generally have a simple Pauli expansion as other observables typicically would, but gives a stringent test for the performance of the method for all observables on the state of interest. The subspace expansion is then performed with $\mathcal{S}^{(l)}$ as expansion operators, and the fidelity $F_L$ of the resulting state is computed with $\ket{\overline{\Psi}}$.  The crossed lines show the performance using fixed projectors at those level, which exactly coincide with the QSE relaxation.  The starred lines show the result of removing $2$ check operators at random and re-performing the QSE expansion to show the performance smoothly interpolates between those limits.  We plot the logical infidelity $1-F_L$ where the physical line denotes the trivial encoding into one qubit and compare the two for a range of values of $p$.  We define the pseudo-threshold to be the value of $p$ for which the logical infidelity in the encoded space is lower than the physical infidelity for the unencoded system, and we see that at both levels there is a pseudo-threshold in this model.  For $S^{(4)}$, the pseudo-threshold is numerically found to be $p=0.50$ for this code and symmetric depolarizing channel.

As an additional test, we examine the logical infidelity as a function of depolarizing noise for the application of a logical $\overline X$ operator which is transversal in this encoding.  The circuit and noise model is depicted in Fig. \ref{fig:transversal_circuit}.  We again see a decisive pseudo-threshold in this case.  We leave open the question of how to best perform non-transversal gates in this model of error mitigation.

\begin{figure}[t]
\centering
\includegraphics[width=8cm]{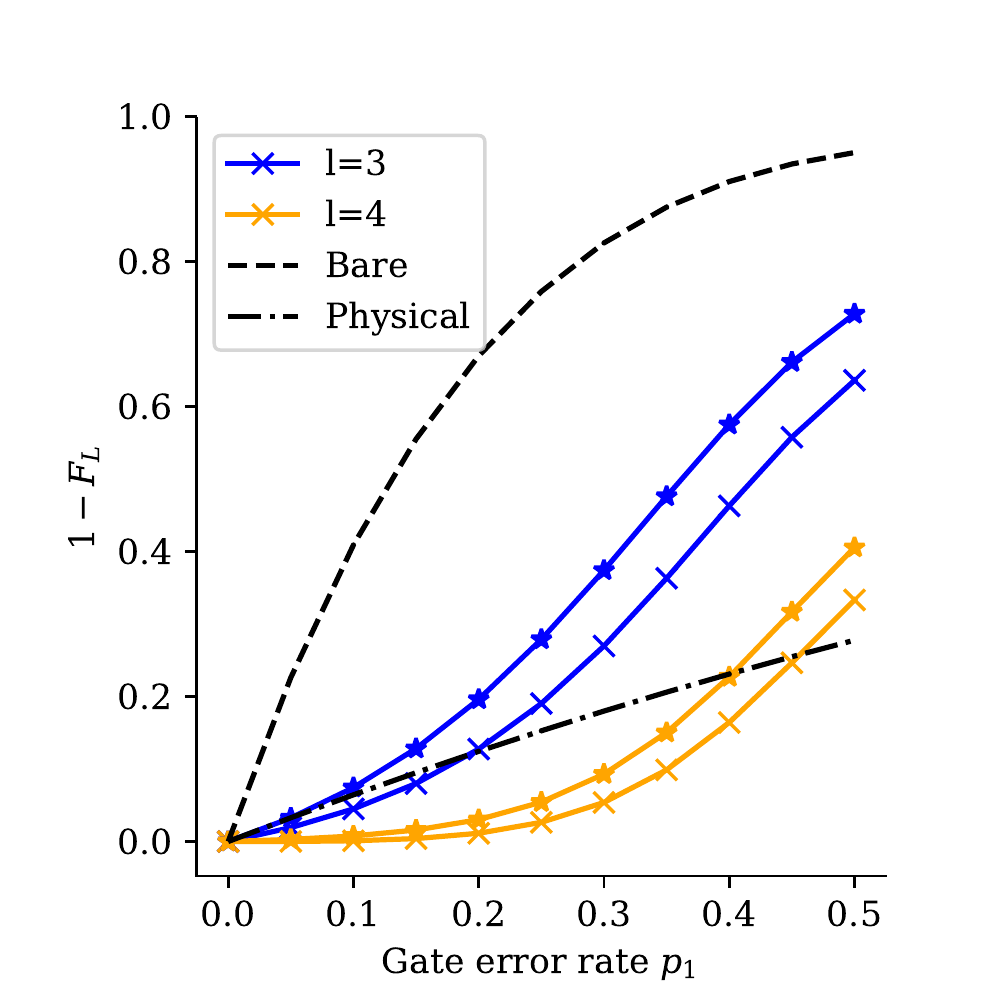}
    \caption{Pseudo-threshold crossover for recovery after applying a logical $X$ gate using the $[[5,1,3]]$ code under an uncorrelated depolarizing channel showing a $p \approx 0.45$ pseudo threshold for the full correction procedure.  We plot the logical infidelity $1-F_L$ where $F_L$ is the logical fidelity for a select state in the code space of the $[[5,1,3]]$ code as a function of the depolarizing probability for each qubit $p$.  The label $l$ denotes the number of products from the stabilizer generators used in the expansion operator set.  The physical line depicts the same error if the logical state is encoded in a single qubit in the standard way.  The ``bare'' line indicates the logical error rate with no recovery procedure applied.  The starred lines close to each level of the hierarchy show an approximation to that level of projection using the QSE projection with $2$ less check operators to demonstrate the smooth performance of the subspace procedure.
    \label{fig:PseudoThresholdTransversalX}}
\end{figure}

\begin{figure}[t]
\centering
\includegraphics[width=8cm]{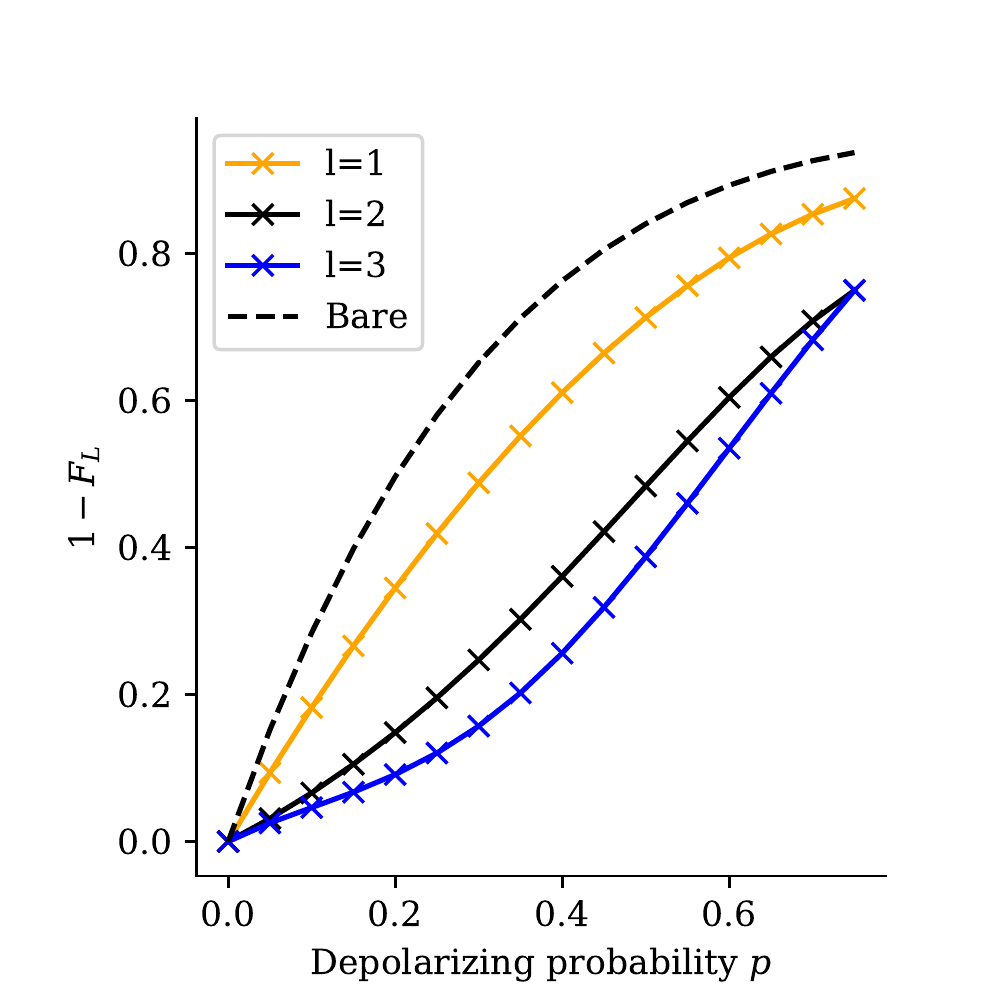}
    \caption{Error as a function of depolarizing probability and included projectors for an H$_2$ molecule at a bond length of $1.50\ \AA$  with a physical encoding.  $F_L$ here is the logical fidelity, which is the same as physical for this encoding.  At this bond length, the ground state wavefunction requires considerable entanglement to be qualitatively correct.  The stabilizer elements used to perform the projection here are the $\alpha$- and $\beta$- number parity as well as the total $X$ operator, which need not be an exact symmetry, given in the Jordan-Wigner encoding as $S=\{ Z_0 Z_2, Z_1 Z_3, X_0 X_1 X_2 X_3 \}$.  The depolarizing channel is applied individually to each qubit on the exact ground state.  As there is no error correction encoding overhead in this case, the expected improvement over the uncorrected solution is always positive.  Hence it is always advantageous in these cases where simple symmetries are known to include these measurements and corrections.
    \label{fig:FermionicPerformance}}
\end{figure}
\subsection{Hydrogen Molecule}
We now look at an example of an unencoded Hamiltonian, which has become a canonical test case for quantum computing in quantum chemistry.  This is the second quantized hydrogen molecule in a minimal, 4 qubit basis.  In the fermionic representation, this has a problem Hamiltonian given by
\begin{align}
    H_p = \sum_{ij} h_{ij} a_i^\dagger a_j + \frac{1}{2} \sum_{ijkl} h_{ijkl} a_i^\dagger a_j^\dagger a_k a_l
\end{align}
where $h_{ij}$ and $h_{ijkl}$ are the one and two electron integrals defined by the basis and geometry of the molecule.  This problem Hamiltonian has a symmetry given by the fermionic number operator $\hat N = \sum_i a_i^\dagger a_i$, and hence the number of fermions in the system is a good quantum number.  As discussed before, however, building projectors directly from the number operator can be cumbersome due to the need to remove all incorrect particle number sectors, and the number of Pauli operators that compose the number operator itself.  

As a result, the $\alpha$ and $\beta$ number parity operators are a more efficient choice of projectors.  In the Jordan-Wigner representation, these are given by $\prod_{i \in \text{even}} Z_i$ and $\prod_{i \in \text{odd}} Z_i$ when an even-odd ordering of orbitals are used.  

The effectiveness of the technique on this system is evaluated numerically by preparing the exact ground state of the hydrogen molecule and subjecting it to an independent depolarizing channel on all $4$ qubits.  We choose as the generating operators $S = \{Z_0 Z_2, Z_1 Z_3, X_0 X_1 X_2 X_3\}$ which up- and down-spin number parity operators as well a non-local operator which need not be an exact symmetry of the Hamiltonian.  The ordering of $l$ matches that given here.  The logical infidelity as a function of the depolarizing probability is plotted in Fig. \ref{fig:FermionicPerformance}.  In contrast to the encoded case, we see an improvement over the whole range of depolarizing strengths.  The stretched geometry of the molecule ensures that a high degree of entanglement is required to achieve a low logical fidelity, making this a sensitive test of performance.  We see that in some cases an improvement of up to $3$x in the logical infidelity.  As the number of operators to measure here is quite modest and the improvement is universal, it suggests this will present an advantageous correction to include in almost all near-term implementations.

\newcommand{\LL}{\Gamma}   
\newcommand{\nsg}{m}  
\newcommand{\nq}{n}   
\newcommand{\PP}{{{\overline P}^{(2)}}}  
\newcommand{\ep}{p} 
\newcommand{\RR}{R}   
\newcommand{\bchi}{{\boldsymbol\chi}}   
\newcommand{\syn}{{s}}   
\newcommand{\bsyn}{{\boldsymbol s}}   

\section{Stochastic sampling for corrections}
\label{sec:random}

As suggested in the exposition on the correction formalism, it is key for efficiency to sample the terms for the correction in a way that reflects the state rather than number of terms.  We outline and analyze a simple stochastic scheme for performing this sampling here.

Suppose that we want to measure the corrected expectation value of some logical operator $\LL$, which can be decomposed into Pauli operators $\Gamma_j$ as $\LL = \tilde \gamma \sum_j \gamma_j \Gamma_j$, where $\sum_j \gamma_j = 1$ and $\gamma_j \geq 0$ from absorbing required signs into $\Gamma_j$.  Projecting this into the code space of some selected code as before, we have
\begin{align}
\mu_\Gamma &= 
\tr \big[\overline P \rho \overline P \LL\big] \notag \\ 
&= \frac{\tilde \gamma}{2^\nsg}\sum_{ij} \gamma_j \tr \big[\rho \LL_j M_i\big]\\
 &= \tilde \gamma \sum_{\bchi \in \{0,1\}^\nsg}\sum_j \frac{\gamma_j}{2^\nsg} \LL_{\bchi, j}\, \label{eq:average_ell}
\end{align}
where $\bchi = (\chi_1,\chi_2,\ldots,\chi_\nsg)$ is a bit string that we use to conveniently enumerate the stabilizer group operators and we define
\begin{align}\label{eq:L_chi}
    \LL_{\bchi, j} =  \tr\big[\rho \LL_j S_{\bchi} \big]\,,\quad S_{\bchi} = S_1^{\chi_1}S_2^{\chi_2}\cdots S_\nsg^{\chi_\nsg}\,
\end{align}
where we note that this also encompasses the measurement of the normalization correction $c$ as well for $\LL=I$.  

To sample the trace stochastically, we may use the coefficients of the terms as a normalized probability distribution.  One may choose the distribution to depend on $\bchi$ as in an importance sampling scheme below, however taking the uniform distribution is perhaps the most straightforward and $p_{\bchi, j} = \frac{\gamma_j}{2^m}$ gives the mean
\begin{align}
    \mathbb{E} \big[ \hat{\mu_\Gamma} \big] &= \tilde \gamma \sum_{\bchi, j} p_{\bchi, j} \mathbb{E}  \left[ \LL_{\bchi, j} \right] \notag \\
    &= \tilde \gamma \sum_{\bchi, j} p_{\bchi, j} (q^{1}_{\bchi, j} - q^{-1}_{\bchi, j}) \notag \\
    &= \tilde \gamma \sum_{\bchi, j, x \in \{-1, 1\}}  x \cdot p_{\bchi, j, x}
\end{align}
where we used $\hat{\mu_\Gamma}$ to emphasize that this is an expected value for our estimator of $\mu_\Gamma$, $q^x_{\bchi, j}$ is the probability of getting a measurement result $x \in \{+1, -1\}$ from measuring the Pauli operator $\Gamma_{\bchi, j}$, and we have lumped this into the probability distribution as $p_{\bchi, j, x} = p_{\bchi, j} q^{x}_{\bchi, j}$.  This has a simple construction for stochastic evaluation, which is to enumerate all the possible terms in the decomposition, draw $N_s$ terms with the probabilities from this distribution which will yield either $+1$ or $-1$ from the Pauli measurements, add them together and divide by $N_s$.  The variance in the estimate will be given by the variance of this estimator divided by $N_s$.  

From our construction, we see that we can view the estimator as a binomial distribution with a probability for the two results $x \in \{-1, +1 \}$ derived from marginalizing over the joint distribution for projector terms $\chi$ and Pauli decomposition terms $j$ to find
\begin{align}
    p_x = \sum_{\chi, j} p_{\chi, j, x}.
\end{align}
As a result, one may write down a particularly simple form of the variance for the estimator given by
\begin{align}\label{eq:variance}
    \Var \big[ \hat{\mu_\Gamma} \big] &= \tilde \gamma^2 p_{+1} (1 - p_{+1}).
\end{align}
To understand how the state influences the variance, we consider a simple example case using the total depolarizing channel and a single Pauli operator $\Gamma$, with $\gamma_i=\tilde \gamma = 1$.  For the total depolarizing channel with probability $w$, we have
\begin{align}\label{eq:total_depolarizing}
  \rho =  (1-w) \ket{\overline \Psi}\!\bra{\overline \Psi} + \frac{w}{2^\nq}\, \openone \,.
\end{align}
In the limit of $w=0$ we have that all states are in the code space, and hence the sum over $\bchi$ trivially collapses, and we have that $p_x = q^x$, which gives the same statistics as the original measurement of $\Gamma$.  Thus in such a case one has no dependence on the number of stabilizer terms used in the expansion.  Hence, adding this procedure to a perfect state is expected to incur no additional cost on average.

Considering the imperfect case $w > 0$, marginalizing over $\chi$ is equivalent to applying the code space projector and hence geometrically analogous to determining the volume of the state in the code space.  This must correspond equate to a portion of the average being $0$, however as we are only capable of measuring $1$ and $-1$, it then must constitute an equal probability of being in $+1$ and $-1$ that is determined by the volume of non-code space the state occupies.  More explicitly
\begin{align}
    \sum_{\boldsymbol\chi} p_{\boldsymbol\chi, j, x} &= \frac{1}{2}(1 + x \tr[\rho \overline P]) p_{j, x}.
\end{align}
As a single Pauli is traceless, in our simple example we find for the case of the totally mixed state that 
\begin{align}
p_{+1} &= \frac{1}{2}(2 - w)q^{+1} \\
\Var \big[ \hat{\mu_\Gamma} \big] &= \frac{1}{4}(2 - w) w (q^{+1})^2
\end{align}
to further simplify, suppose we were measuring the $+1$ eigenstate of $\Gamma$, so that $q^{+1} = 1$, then
\begin{align}
\Var \big[ \hat{\mu_\Gamma} \big] &= \frac{1}{4}(2-w)w
\end{align}
then we see as expected, that for a perfect state $(w=0)$ the variance is minimal and independent of the number of terms, and that the variance increases as the state quality degrades.

The general picture of viewing the sum over $\bchi$ as reflecting the volume of space attached to the projector lets one easily reason about the generalization of this scheme to sampling with recovery.  In that case, we simply attach one more probability which allows us to sample over the different selected errors $E_i$, with probability $p_{E_i}$ and proceed as before.  The key difference is that we see the marginalization over $\bchi$ is now better conditioned, as it is determined by the ratio of the volume of recovered space to the volume of Hilbert space rather than the volume of the code space to the volume of Hilbert space.  As a result, the sample variance for recovery may be lower than the sample variance for strict projection.  However as discussed, the ultimate quality of recovery is expected to be superior for strict projection due to removing errors up to weight $d-1$ instead of $(d - 1) / 2$.

One potential way to suppress the numbers of samples required is to use variance reduction techniques such as importance sampling.  This approach requires {\it a priori} knowledge of the values of $\LL_{\bchi, j, x}$, and samples the high weight terms preferentially while applying a correction to the measured values to remain unbiased.  One approach is to sample the bit string $\bchi$ according to its Pauli-weight $W_{\bchi}$, i.e, the number of qubits that the stabilizer operator acts non-trivially on.  For the single-qubit depolarizing channel with error probability $p$, one may sample the bit string $\bchi$ with probability proportional to $(1-p)^{W_{\bchi}}$.  This is based on the intuition that the quantum information stored in low-weight operators decays more slowly under local noisy channels. 

To be more explicit in the construction for the random sampling method discussed here applied to the recovery procedure discussed in Section~\ref{sec:recovery_operations}.  
Suppose that the projector on the subspace corresponding to the error operator $E_\alpha$ is
\begin{align}
    \overline P_\alpha = \prod_j  \frac{1}{2}\Big(1 + (-1)^{\syn_{\alpha, j}} S_j\Big)\,,
\end{align}
where $\syn_{\alpha, j} = 0, 1$ and $\bsyn_\alpha = (\syn_{\alpha, 1},\syn_{\alpha, 2},\ldots, \syn_{\alpha, \nsg})$ are the syndromes of the $\alpha$-th error.  The recovered state takes the form
\begin{align}
  \mathcal R (\rho) &= \sum_\alpha \RR_\alpha \overline P_\alpha \rho \overline P_\alpha \RR_\alpha^\dagger\\
  &= \overline P \Big(\sum_\alpha \RR_\alpha  \rho  \RR_\alpha^\dagger \Big)\overline P\,.
\end{align}
The expectation value for the recovered state reads
\begin{align}
 \tr\big[\mathcal R (\rho)\LL\big] &= \sum_{\alpha, j} \tr \big( \RR_\alpha \rho \RR_\alpha^\dagger \overline P \LL_j \overline P\big)\\ 
 &= \frac{1}{2^\nsg}\sum_{\alpha, \bchi, j} \tr \big(\RR_\alpha\rho \RR_\alpha^\dagger \LL_j S_{\bchi} \big)
\end{align}
where $\LL_\alpha = \RR_\alpha^\dagger \LL \RR_\alpha$ is a logical operator.  Hence if we absorb the sign into the logical operator through a careful choice of recovery operator, we may generalize the stochastic sampling scheme to taking the expectation value as
\begin{align}
  \mathbb{E} \big[ \hat{\mu_\Gamma} \big] &= \tilde \gamma \sum_{\bchi, j, \alpha, x \in \{-1, 1\}} x \cdot p_{\bchi, j, \alpha, x}
\end{align}
where for uniform sampling we choose
\begin{align}
  p_{\bchi, j, \alpha} = \frac{\gamma_i b_\alpha}{2^m}
\end{align}
with $\sum_\alpha b_\alpha = 1$ and $b_\alpha > 0$ by choice of the recovery operators.  In this case, the stochastic sampling algorithm is given by choosing a Pauli operator defined by $\RR_\alpha^\dagger \LL_j S_{\bchi} \RR_\alpha$ with probability $p_{\bchi, j, \alpha}$, recording the series of $+1$, $-1$ results, and finding their average as before.  The calculation of the variance follows as in the previous case.  To reduce variance, the frequency of sampling $\alpha$ maybe chosen to be proportional to the error probability of $E_\alpha$ for the expected errors on the physical system of interest.

\section{Discussion}
As has been conjectured before, one of the best uses of early quantum devices may be to tune quantum error correcting codes under actual device conditions~\cite{Iyer:2018}.  The modeling of true noise within the device is incredibly difficult as the system size grows, and studying which codes excel under natural conditions and how to optimize them may lead to progress towards fully fault tolerant computation.  Indeed, knowledge of biased noise sources can vastly increase the threshold of a given code~\cite{Tuckett:2018}.  The tool we have provided here gives a method to experimentally study the encoding through post-processing while removing the complication of fault-tolerant syndrome measurement or fast feedback.  This allows one to explore a wider variety of codes experimentally before worrying about these final details.  We propose that one can run simple gate sequences in the logical space with known results, and use the post processing decoder here to study the decay of errors as stabilizers are added.  This limit will inform the propagation of logical errors in the system and allow one to make code optimizations before full fault tolerant protocols are available. 

We note that this type of decoder benefits greatly from the fact that check measurements need not be geometrically local to be implementable in a realistic setting.  This allows one to explore and utilize codes that are not geometrically local on the architecture in use, which may have nicer properties with respect to distance and rate than geometrically local codes.   Moreover, they naturally allow implementation of recent fermionic based codes, such as Majorana loop stabilizer codes~\cite{Jiang2018} or variations of Bravyi-Kitaev superfast~\cite{Setia:2018} thought to be good candidates for near-term simulations, without the need for complicated decoding circuits or ancilla for syndrome measurements.

Moreover, as this method is a post-processing method, it is entirely compatible with the extrapolation techniques introduced for error mitigation~\cite{Temme2017error,Endo:2017,Otten:2018}.  In these techniques, one artificially introduces additional noise to extrapolate to a lower noise limit.  To apply this technique to quantum subspace expansions, one simply needs to perform the extrapolation on each of the desired matrix elements, then proceed as normal. 

In this work, we have introduced a method for mitigating errors and studying error correcting codes using a post processing technique based on quantum subspace expansions.  We showed that in implementations of this method, one can achieve a pseudo-threshold of $p \approx 0.50$ under a depolarizing channel acting on single qubits in the $[[5,1,3]]$ code and made connections to the traditional theory of stabilizer codes.  We believe this method has the potential to play a role in the development and optimization of quantum codes under realistic noise conditions as well as the ability to remove errors from early application initiatives.

\section{Acknowledgements}
We are grateful for input and helpful discussions with Kenneth Brown, Fernando Brandao, and Dave Bacon on this manuscript.
\bibliographystyle{apsrev4-1_with_title}
\bibliography{references}
\end{document}